\magnification\magstep1
\tolerance=1600
\parskip = 6pt
\def\dt{{\cdot}}
\def\half{{1\over 2}}
\def\hhalf{\textstyle{1\over 2}}
\def\ref#1{$^{[#1]}$}
\def\pagenumber{\footline={\hss\tenrm\folio\hss}}

\def\half{{\textstyle{1\over 2}}}

\def\tr{\hbox{tr}_\alpha}

\def\ze{{\bf Z}}

\def\ag{{g}}

\def\dt{{\cdot}}

\def\ref#1{$^{[#1]}$}

\def\pagenumber{\footline={\hss\tenrm\folio\hss}}
\nopagenumbers
\def\ag{\alpha}
\def\bg{\beta}
\def\gg{\gamma}

\def\ih{{\hat \imath}}
\def\jh{{\hat \jmath}}

\def\Thf #1. #2.{\Theta\left[{\textstyle{#1\atop #2}}\right]}
\def\Chf #1. #2.{\bar\Theta\left[{\textstyle{#1\atop #2}}\right]}
\def\ZZ #1. #2.{Z_k\left[{\textstyle{#1\atop #2}}\right]}

\nopagenumbers
\rightline{hep-th/9711114}
\rightline{IFP/97-21/UNC}
\vskip 20pt
\footline={\sevenrm
\hfil$^\ast$ Supported in part by the U.S. Department of
Energy under Grant No. DE-FG 05-85ER40219/Task A\hfil}
\vskip 3pt
\centerline{\bf PARTITION FUNCTIONS, DUALITY AND THE TUBE METRIC   
$^\ast$}
\vskip 50 pt
\centerline {\bf L. Dolan$^1$ and M. Langham$^2$}
\vskip 12pt
\centerline{\it
Department of Physics and Astronomy,University of North Carolina}
\centerline{\it Chapel Hill, North Carolina 27599-3255, USA}
\vskip 50pt
\centerline {\bf ABSTRACT}
{\rightskip=18 true mm \leftskip=18 true mm  \noindent

The partition function of type IIA and B strings on $R^6 \times K3$,
in the {\bf $T^4/Z_2$} orbifold limit, is explicitly computed 
as a modular invariant sum over spin strutures
required by perturbative unitarity in order to extend the analysis
to include type II strings on $R^6 \times W4$, where $W4$ is associated with
the tube metric conformal field theory, given by the degrees of freedom
transverse to the Neveu-Schwarz fivebrane solution. This generates
partition functions and perturbative spectra of string theories in 
six space-time dimensions, associated with the 
modular invariants of the level $k$ affine $SU(2)$ Kac-Moody algebra. 
These theories provide a conformal field theory
(i.e. perturbative) probe of non-perturbative (fivebrane) vacua.  
We contrast them with theories whose 
$N=(4,4)$ sigma-model action contains $n_H=k +2$ hypermultiplets 
as well as vector supermultiplets, and where $k$ is the level just mentioned. 
In Appendix B we also give a $D=6$, $N=(1,1)$ `free fermion' string model
which has a different moduli space of vacua from the 81 parameter
space relevant to the above examples.
}
\vfill
\centerline{November 1997}
\vskip80pt
\eject

\pagenumber
\centerline{\bf1. Introduction}
\vskip4pt
The non-perturbative formulation of string theory is motivated by dualities,
but ultimately it should provide the identity of 
a set of degrees of freedom on which the theory is consistently defined. 
 
In perturbative string theory, the GSO projections lead to modular invariance
of the one-loop amplitudes. We further recall\ref{1-3}
that in string perturbation theory
there are requirements of space-time factorization
for vertices and one-loop modular invariance which guarantee higher-loop
modular invariance. These requirements are related to the conservation 
at vertices of quantum numbers associated with the projections,
and are satisfied by  models that are unitary on a set of states 
picked out from the whole Fock space by a set of 
projections defined by number operators. 
In perturbative string theory,
the projections in the various sectors are a primary concept.

Currently, the backbone of non-perturbative string theory is 
a web of various string perturbative expansions, described by dual pairs
of theories\ref{4,5} 
and more generally by matrix theory\ref{6}. 
To construct a complete non-perturbative formulation, 
the guiding principle which replaces
perturbative unitarity is that strong coupling (non-perturbative) extensions
of existing perturbative string models appear to be well defined in terms of 
other perturbative models. In particular, a dual pair of string theories
have the same moduli space of vacua, which is already present in the low
energy (field theory) description of either theory. In this way 
the moduli spaces (different ones correspond to different compactifications)  
become a primary concept. 

In this paper, we analyze several superstring vacua in some detail in order
to extract from perturbative formulations relevant non-perturbative information
which may be useful in shaping the appropriate set of 
states on which the theory is defined. 

In sect.2, we consider the type IIA and IIB superstrings on 
$R^6 \times K3$ and give the partition function explicitly as sum over
spin structures for its orbifold limit ${\bf R}^6\times {{\bf T}^4/{\bf Z}_2}$.
We identify the T-duality group for this restricted class of $K3$ 
compactifications, and show that it is an invariance of the partition function,
although its status as an invariance of the full non-perturbative spectrum
of this theory remains conjectural. 
The generalized GSO projections that define these models are then used
in sect.3 to derive the massless spectra, which are seen to correspond to 
$D=6$ theories with space-time supersymmetries $N=(1,1)$ and $N=(2,0)$  
respectively that occur for generic $R^6 \times K3$ compactifications. 
In sect.4 analogous formulae are given for
the type II superstrings on ${\bf R}^6\times W4/{\bf Z}_2$,
where $W4$ is the `tube metric' conformal field theory (cft) associated
with the transverse degrees of freedom of the type II NS fivebrane\ref{7-10}.
In sect.5, we recall 
the relationships\ref{11-16} among this cft, points in the moduli space
corresponding to non-abelian (enhanced) gauge symmetry, the 
$D=2$, $N=(4,4)$ supersymmetric 
theories with vector and $n_H = k+2$ hypermultiplets,
level $k$ $\widehat{SU}(2)$ affine algebras, 
and the theory of $n_H$ coincident NS fivebranes. We discuss the norms of
both continuous and discrete states in the `tube metric' cft, and a 
connection to exact results on Liouville cft\ref{17}.
Conclusions and comments are found in sect.6 which discusses   
the presence of enhanced gauge symmetry 
and its incorporation in type II variables. 
We also give a type II free fermion string model with conventional (NS-NS)
non-abelian symmetry
(for the gauge group $SU(2)^4$) which has a different moduli space
of vacua  from the above examples. This model shows that all known compactifications
of string theory with $N=(1,1)$ supersymmetry in $D=6$ are not on the
same moduli space of vacua. 
  
\vskip15pt
\centerline{\bf 2. Partition function}
\vskip5pt
We first compute explicitly
type II strings on ${\bf R}^6\times {{\bf T}^4/{\bf Z}_2}$,
where ${\bf T}^4/{\bf Z}_2$ is the ${\bf Z}_2$ orbifold limit of $K3$.
In the light-cone description, the left and right-moving
modes are each
taken to be described by 8 bosonic and 8 fermionic worldsheet (primary) 
fields:
$\tilde A^\ih_n, \tilde A_s^I, \tilde \psi_r^\ih, \tilde \psi_r^I$;
$A^\ih_n,  A_s^I, \psi_r^\ih, \psi_r^I$; $1\leq \ih,\leq 4$ and
$5\leq I,\leq 8$, where the
superscript $\ih$ refers to the transverse spatial degrees of freedom, 
the superscript $I$ to the internal ones, and the subscripts $s,r$ each
correspond to either integer or half-integer modding depending on the sector. 
The partition function or one loop contribution to the vacuum to vacuum 
amplitude in $D$ space-time dimensions is
$$\Lambda = -\,{1\over {4\pi {(\alpha')^{D\over 2}}}} \int_{\cal F}  
d^2 \tau (\hbox{\rm Im}\,\tau)^{-2-({{D-2}\over 2})}  	
\,|f(\omega)|^{-2(D-2)} \,\Lambda_f\eqno(2.1)$$
where $\Lambda_f$ is the partition function for the
fermionic and internal bosonic degrees of freedom,
$$\Lambda_f = \sum_{\alpha \in \Omega} \delta_\alpha \,\tr
\{\bar \omega^{\tilde L_0-\half} \omega^{L_0 - \half} \prod_{\beta
\in\Omega}
P_{\alpha, \beta}\},\eqno(2.2)$$
{\it i.e.} the spectrum of a theory will consist of a set of sectors
$\Omega$,
characterized by the modding of the internal bosons ($s\in \ze$, untwisted),
($s\in \ze+\half$, twisted), and of the fermions ($r\in\ze$, untwisted Ramond (R)
or twisted Neveu-Schwarz (NS)), and
($r\in \ze+\half$, untwisted NS or twisted R).  
The quantities $\delta_\alpha$ and the projection operators
$P_{\alpha,\beta}$ are discussed below.  The integration region
${\cal F}=\{\tau\,:\, |\tau|>1\; 
|\hbox{\it Re}\tau |<\half\}$ is a fundamental region for the modular
group that is generated by $\tau \rightarrow \tau + 1,\;
\tau \rightarrow -{1\over\tau}$; 
and $\tilde L_0$, $L_0$ refer to left, right movers. 

To define the orbifold choose a complex basis for
the internal fermions, for example for the left-movers: 
$f^1={\scriptstyle{1\over{\sqrt 2}}} 
(h^5 + i h^6)$, 
$\bar f^1={\scriptstyle{1\over{\sqrt 2}}} (h^5 - i h^6)$,
$f^2={\scriptstyle{1\over{\sqrt 2}}} (h^7 + i h^8)$,
$\bar f^2={\scriptstyle{1\over{\sqrt 2}}} (h^7 - i h^8)$.
Then the ${\bf Z}_2$ transformation $\theta$ acting on the internal 
fermions in terms of the number operator 
$F = \sum_{i=1,2; r} :f_r^i \bar f_{-r}^i:$ $\,$ (where
$:f_0^i \bar f_0^j: = - :\bar f_0^j f_0^i:$), so that
$\theta = (-1)^F$;  and similarly $\theta$ acts on the 
internal bosons by $\theta  A_s^I \theta^{-1} = -  A_s^I$.
Oscillators with space-time indices are invariant under $\theta$,
and $D=6$. 

A sector $\alpha$ is labelled by a twelve-dimensional vector 
whose components are 0 for NS and 1 for R:
$$\rho_\ag = (\tilde\rho_1,\ldots,\tilde \rho_4;\tilde\rho'_1,
\tilde\rho'_2;\rho_1,\ldots,\rho_{4};\rho'_1,\rho'_2)\,  
\eqno(2.3)$$ 
This vector corresponds to boundary conditions of
left- and right- modes separately described by 4 real
and 2 complex fermions.         

The set of states on which the theory is unitary is 
specified by states that survive projections defined by number
operators which generalize
the GSO projection. The projections are defined by requiring the
parity of the number
operators, $N_\beta$ defined in (2.9), 
to take on definite values
$\epsilon(\alpha, \beta)$ on any state in
the sector $\alpha$, {\it i.e.}
$$(-1)^{N_\beta}\bigr|_\ag=\epsilon(\alpha,\beta) ,\eqno(2.4)$$
where each $\epsilon(\alpha, \beta)$ is either $\pm 1$.  
The (perturbative) spectrum
of a model
is specified by a set of sectors $\Omega$, 
together with a set $\{(-1)^{N_\beta}\, : \, \beta \in
\Omega\}$
of parity operators, and their values $\epsilon(\alpha, \beta)$
on the
sectors $\alpha\in\Omega$.  

Eq. (2.2) can be expressed as a sum over spin
strutures\ref{1}:
$$\Lambda_f =
{1\over{2^{K+1}}}
\sum_{\alpha \in \Omega}
\sum_{\bg\in\Omega} \delta_\ag \; \epsilon(\ag, \bg)\; \tr
\{\bar \omega^{\tilde L_0-\half} \omega^{L_0 - \half}
(-1)^{N_\bg}\}\, 
\eqno(2.5) $$  
\noindent 
where K is the number of basis vectors which generate $\Omega$.
We denote the trace in eq.(2.5) by
$\{\ag,\bg\}$, so that, without the factor of $2^{-K-1}$,
the sum is 
$$\sum_{\ag,\bg} \delta_\ag \; \epsilon(\ag, \bg) \;
\{\ag,\bg\}\, , \eqno(2.6)$$
where
$$\eqalignno{\{\ag,\bg\}\ &= \tr
\{\bar \omega^{\tilde L_0-\half} \omega^{L_0 - \half}
(-1)^{N_\bg}\}\cr
&= |\omega|^{-1} |f(\omega)|^{-20} \prod_{i=1}^{4}
\left({\Chf \tilde\rho_\alpha^i. \tilde\mu_\beta^i.
(0|\tau)}\right)^{1\over2} \prod_{i=5}^{6}
\left({\Chf \tilde\rho_\alpha^i. \tilde\mu_\beta^i.
(0|\tau)}\right)\cr & \;
\hskip 3pt\times\prod_{j=1}^{4}
\left({\Thf \rho_\alpha^j. \mu_\beta^j.(0|\tau)}\right)^{1\over2}
\prod_{j=5}^{6} \left({\Thf \rho_\alpha^j. \mu_\beta^j.(0|\tau)}\right)
\hskip10pt\times {\rm internal \,bosons}
\, , &(2.7)}$$
and $(\tilde\rho^i_\alpha, \rho^j_\alpha) $
and $(\tilde\mu^i_\beta, \mu^j_\beta)$ are the 
the twelve-component vectors describing the sectors $\alpha$ and
$\beta$ respectively, {\it i.e.} the components are 0 for NS and 1 for
R [see for example (2.8)].  ${\Thf \rho^i. \mu^i.(0|\tau)}$
and $f(\omega)$ are given by (2.14), and
$\delta_\ag = \delta_\ag^L\delta_\ag^R$ where  
$\delta_\ag=1$ if the states of the sector $\ag$
are space-time bosons and  $\delta_\ag=-1$ if the states are space-time
fermions. 
A consistent (perturbative) string theory is such that under
modular transformations 
the integrand of (2.1) is invariant.  

The type II string on ${\bf R}^6\times {{\bf T}^4/{\bf Z}_2}$ has 
eight sectors, whose fermion boundary condition vectors (2.3) are given by
$$\eqalignno{\rho_{b_2}&= (1^4,1^2; 0^4,0^2)\hskip 40pt
\rho_{b_0b_1b_2}= (1^4,0^2; 0^4,1^2)\cr
\rho_{b_0b_2}&= (0^4,0^2; 1^4,1^2)\hskip 47pt
\rho_{b_1b_2}= (0^4,1^2; 1^4,0^2)\cr
\rho_{b_0b_1}&= (0^4,1^2; 0^4,1^2)\hskip 55pt
\rho_{b_1}= (1^4,0^2; 1^4,0^2)\cr
\rho_{b_0}&= (1^4,1^2; 1^4,1^2)\hskip 57pt
\rho_{\phi}= (0^4,0^2; 0^4,0^2) \,, &(2.8)}$$ 
where $\{\phi, b_0, b_0b_2, b_2\}$ are the sectors 
that have untwisted bosons, 
and the ${\bf Z}_2$ twisted sectors 
are written as
$\{b_0b_1, b_1, b_1b_2, b_0b_1b_2\}$.
For this theory, the eigenvalues $\epsilon (\alpha,\beta)$ 
of the parity operators are given in Table 1, where
$\lambda,\rho,\mu,\nu$ take values $\pm1$, and different choices
of $\lambda,\rho,\mu,\nu$ do not change the theory. 
Table 1 is derived by requiring 
modular invariance for the part of $\Lambda$ given by
${1\over 8}
\sum_{\alpha,\bg \in \{\phi, b_0, b_0b_2, b_2\}}
\delta_\ag \; \epsilon(\ag, \bg)\; \tr
\{\bar \omega^{\tilde L_0-\half} \omega^{L_0 - \half}
(-1)^{N_\bg}\}$,
and then computing the remaining values of $\epsilon(\ag,\bg)$
using $\epsilon(\ag, \bg\gg) = \epsilon(\ag, \bg)\;
\epsilon(\ag,\gg)\,$, which follows from (2.4).

\vskip 12pt

\hskip-40pt\hfil\hbox{\raise 25mm\vbox{\hbox{$\ag$}\hbox{\big\downarrow}}
\vbox{\offinterlineskip
\halign{\vrule#&
  \strut\quad\hfil#\quad&\vrule#&
  \strut\quad\hfil#\quad&
  \strut\quad\hfil#\quad&
  \strut\quad\hfil#\quad&
  \strut\quad\hfil#\quad&
  \strut\quad\hfil#\quad&
  \strut\quad\hfil#\quad&
  \strut\quad\hfil#\quad&
  \strut\quad\hfil#\quad&\vrule#\cr
\multispan{12} \hfil $\bg \longrightarrow $\hfil\cr
\noalign{\vskip 4pt\hrule}  
height2pt&\omit&&\omit&\omit&\omit&\omit&
\omit&\omit&\omit&\omit&\cr
&$\epsilon(\ag,\bg)$\hfil&&\O&$b_0$&$b_1$&$b_2$&
$b_0b_1$&$b_0b_2$&$b_1b_2$&$b_0b_1b_2$&\cr
height2pt&\omit&&\omit&\omit&\omit&\omit&
\omit&\omit&\omit&\omit&\cr
\noalign{\hrule}
height2pt&\omit&&\omit&\omit&\omit&\omit&
\omit&\omit&\omit&\omit&\cr
&\O&&1&1&$1$&$-1$&$1$&$-1$&$-1$&$-1$&\cr
&$b_0$&&1&$\lambda$&$\rho$&$\mu$&$\lambda\rho$&$
\lambda\mu$&
$\rho\mu$&$\lambda\rho\mu$&\cr
&$b_1$&&1&$\rho$&$\rho$&$\nu$&$1$&$\rho\nu$&
$\rho\nu$&$\nu$&\cr
&$b_2$&&1&$-\mu$&$\nu$&$\mu$&$-\mu\nu$&$-1$& 
$\mu\nu$&$-\nu$&\cr
&$b_0b_1$&&1&$\lambda\rho$&$1$&$-\mu\nu$&
$\lambda\rho$&$-\lambda\rho\mu\nu$&
$-\mu\nu$&$-\lambda\rho\mu\nu$&\cr
&$b_0b_2$&&1&$-\lambda\mu$&$\rho\nu$&$-1$&$-\lambda\rho\mu\nu$&
$\lambda\mu$&
$-\rho\nu$&$\lambda\rho\mu\nu$&\cr
&$b_1b_2$&&1&$-\rho\mu$&$\rho\nu$&$-\mu\nu$&$-\mu\nu$&$\rho\nu$&
$-\mu\rho$&$1$&\cr
&$b_0b_1b_2$&&1&$-\lambda\rho\mu$&$\nu$&$\nu$&$-\lambda\rho\mu\nu$&
$-\lambda\rho\mu\nu$&
$1$&$-\lambda\rho\mu$&\cr
height2pt&\omit&&\omit&\omit&\omit&\omit&
\omit&\omit&\omit&\omit&\cr
\noalign{\hrule\vskip8pt }
\multispan{12} \hfil {\sl Table 1}\hfil\cr
}}}\hfil    
\vskip6pt 
In the fermionic picture, we define the parity of the number operator
$N_\beta$ acting on the sector $\alpha$ by
$$(-1)^{N_\beta}\bigr|_\alpha = (-1)^{\rho_\beta\dt F} \bigr|_\alpha\, 
,\eqno(2.9)$$
where $F$ is a vector
whose components are the operators $ F_j = \sum_{r}:
f_r^j \bar f_{-r}^j:$
for complex fermions and $\sum_{r\ge 0}^\infty \psi_{-r}^j\psi_r^j$ 
for real fermions, and $r$ is modded according to the boundary condition 
of the ``$j^{th}$''fermion in the sector $\alpha$.
$$\rho_\bg\dt F = \sum_{j=1}^4\tilde\rho_j \tilde F_j
+ \sum_{j=1}^2\tilde\rho'_j \tilde F'_j
+ \sum_{j=1}^{4}\rho_j F_j
+ \sum_{j=1}^{2}\rho'_j F'_j
\, \eqno(2.10)$$
and the sums $\tilde F_j$ and $F_j$ distinquish left and right movers,
but the pair $\bar f$ and $f$ denotes a complex fermion
which is either wholly left moving or right moving. 
In general, the projection operators are defined by
$$ P_{\ag, \bg} = \half \{1 + \epsilon(\ag,\bg)
(-1)^{N_\bg}\},\qquad \bg\in
\Omega\,.\eqno(2.11)$$ 
Since $(-1)^{N_{\bg\gg}} = (-1)^{N_\bg}
(-1)^{N_{\gg}}$, it follows that 
$$\prod_{\bg\in\Omega} P_{\ag, \bg} = \prod_{i=0}^{K+1}
P_{\ag,\beta_i}\, \eqno(2.12)$$
where, for this theory $K=2$ and $\beta_0 = b_2, \beta_1 = b_0b_2,
\beta_2 = b_0b_1$.  
The conventional GSO projections involve only fermions:
$$\eqalignno{P_{\ag, b_2 } &= \half \{1 + \epsilon(\ag,b_2)
(-1)^{\rho_{b_2}\dt F}\}&(2.13a)\cr
P_{\ag, b_0b_2 } &= \half \{1 + \epsilon(\ag,b_0b_2)
(-1)^{\rho_{b_0b_2}\dt F}\}\,&(2.13b)}$$
while the ${\bf Z}_2$ projection is incorporated in
$$\eqalignno{P_{\ag, b_0b_1 } &= \half \{1 + \epsilon(\ag,b_0b_1)
\theta\}\cr
\theta &= (-1)^{(\rho_{b_0b_1}\dt F) + N}\cr
N &=\sum_{r>0} {\textstyle{1\over r}} (\tilde A_{-r}^I
\tilde A_{r}^I + A_{-r}^I A_{r}^I ) \,&(2.13c)}$$ 

The functions in (2.7) have $\omega = e^{2\pi i \tau}$ and
$${\Thf \rho. \mu.(\nu|\tau)} = \sum_{n\in\ze} e^{i\pi\tau(n
+{\rho\over 2})^2}
e^{i2\pi(n+{\rho\over 2})(\nu+{\mu\over 2})
-i{\pi\over2}\rho\mu }\,  \eqno(2.14a)$$
$${\Chf \tilde\rho. \tilde\mu.(\nu|\tau)} = \sum_{n\in\ze}
e^{-i\pi\bar\tau(n
+{\tilde\rho\over 2})^2}
e^{-i2\pi(n+{\tilde\rho\over 2})(\nu+{\tilde\mu\over 2})
+i{\pi\over2}\tilde\rho\tilde\mu }\, \eqno(2.14b)$$
$$\eta (\tau) = \omega^{1\over 24} f(\omega) 
= \omega^{1\over 24}  \prod_{n=1}^\infty (1 - \omega^n)\, . \eqno(2.14c)$$ 
The partition function for  
type II strings on ${\bf R}^6\times {{\bf T}^4/{\bf Z}_2}$ is thus
specified by eq.'s (2.8),(2.13) and Table 1. Using this data in (2.1,2,6),
we find that many of the spin structures give zero contribution; 
but also that modular invariance of the integrand of (2.1) can be checked 
explicitly using the $\epsilon(\alpha,\beta)$ eigenvalues in Table 1
and the modular transformation properties of the functions in (2.14). 
This assures that one-loop n-point functions as well as the 
0-point function are modular invariant. 

Collecting the contributions from the different (non-zero) spin structures,
we have for type II strings on ${\bf R}^6\times {{\bf T}^4/{\bf Z}_2}$ that
$$\eqalignno{\Lambda &= -\,{1\over {4\pi {(\alpha')^3}}}\int_{\cal F}
d^2 \tau (\hbox{\rm Im}\,\tau)^{-4}
\,|\eta(\tau)|^{-8} \,\Lambda'_f&(2.15a)\cr
\Lambda'_f &= {\textstyle 1\over 8} 
[\, {\theta_{\Gamma_{4,4}} (\bar\tau,\tau)\over |\eta (\tau)|^8 }
\quad |\eta (\tau)|^{-8} (\bar\theta_3^4 - \bar\theta_4^4 
- \bar\theta_2^4 )  (\theta_3^4 - \theta_4^4 - \theta_2^4)\cr 
&\hskip20pt +  {2^4 |\eta (\tau)|^4 \over \bar\theta_2^2 \theta_2^2}
\quad |\eta (\tau)|^{-8} \,\bar\theta_3^2 \bar\theta_4^2 \theta_3^2
\theta_4^2 \,( 1 - 1 - 1 + 1)\cr
&\hskip20pt +  {2^4 |\eta (\tau)|^4 \over \bar\theta_4^2 \theta_4^2}
\quad |\eta (\tau)|^{-8} \,\bar\theta_3^2 \bar\theta_2^2 \theta_3^2
\theta_2^2 \,(2 - 2)\cr    
&\hskip20pt +  {2^4 |\eta (\tau)|^4 \over \bar\theta_3^2 \theta_3^2}
\quad |\eta (\tau)|^{-8} \,\bar\theta_4^2 \bar\theta_2^2 \theta_4^2
\theta_2^2 \,(2 - 2)\,\,]\,.
&(2.15b)}$$ 
In (2.15b), the factors $(1 - 1 - 1 + 1)$ are all space-time boson 
contributions, while in lines 3 and 4 each the factors $(2 -2)$  
contribute 2 from space-time bosons and (-2) from space-time fermions. 
In line 1, the lattice theta function
$\theta_{\Gamma_{4,4}}(\bar\tau,\tau) = \sum_{p_L,p_R\in \Gamma_{4,4}}
\bar\omega^{\half p_L^2} \omega^{\half p_R^2}$
is defined for any even, self-dual eight-dimensional
Lorentzian lattice $\Gamma_{4,4}$. 
Using the Poisson resummation formula,
one can show the quantity ${\theta_{\Gamma_{4,4}} (\bar\tau,\tau)
\over |\eta (\tau)|^{8}}$ is
modular invariant.  
There is a 16-parameter family of such lattices 
generated by an 
$0(4,4)$ transformation of the $\oplus 4P_2$ lattice, where 
$P_2$ is a two-dimensional lattice with signature $(+-)$ 
defined by basis vectors $\alpha_1^I = {\textstyle{1\over\sqrt 2}}
(1,1);\, \alpha_2^I = {\textstyle{1\over\sqrt 2}} 
(-1,1)$ with metric $\left(\matrix{1&0\cr 0&-1\cr}\right)$.
Since the $D=6$ mass squared is invariant under the $0(4)\times 0(4)$
subgroup of the non-compact group $0(4,4)$, 
the number of different parameter families of 
string models is given by the dimension of the coset
${0(4,4)\over{0(4)\times 0(4)}}$ which is 16. 
The discrete group $0(4,4;{\bf Z})$ leaves invariant the lattice theta
function $\theta_{\Gamma_{4,4}}(\bar\tau,\tau)$
for any given member of the 16-parameter family of lattices.
Therefore the partition function given in (2.15) of the type II string
on ${\bf R}^6\times {{\bf T}^4/{\bf Z}_2}$
has an explicit T-duality symmetry group
$$G_T =  0(4,4;{\bf Z})\eqno(2.16)$$ 
which is an invariance of the perturbative spectrum.
In (2.15) we use $\theta_3 = {\Thf 0. 0.(0|\tau)};\,
\theta_4 = {\Thf 0. 1.(0|\tau)};\,
\theta_2 = {\Thf 1. 0.(0|\tau)};\,
{\Thf 1. 1.(0|\tau)} = 0$.
The partition function for type II strings on ${\bf R}^6\times {\bf T}^4$
is given in Appendix A for later comparisons.
 
\vskip15pt
\centerline{\bf3. Massless and lowest lying massive perturbative spectra}
\vskip5pt
In this section we work out the massless and lowest lying massive spectra
in these theories, using the projection operators (2.13), rather than 
directly using facts 
about the cohomology and moduli space of K3. 
Our procedure insures that the spectrum at a given mass level agrees with
the coefficient of a suitable power of $\bar\omega\omega$ in (2.15). 
The partition function (2.15) describes a precise set of states on
which the theory is known to satisfy (perturbative) unitarity. 

String theories that include both perturbative and non-perturbative states
(such as type IIA on ${\bf R}^6\times K3$ at values of the $K3$ moduli 
corresponding to enhanced non-abelian massless gauge bosons)
are not yet sufficiently defined to allow identification of a consistent 
set of states analogous to the perturbative situation. 
Nonetheless it is conjectured that such a set of states
exists on which unitarity for both perturbative and 
non-perturbative states, and S-duality could be proved. 
In this framework, the massless spectrum of the target space theory
is frequently computed either via a field theory compactification of
type IIA supergravity on ${\bf R}^6\times K3$, using the Betti numbers
of $K3$, or from the moduli space of the 2-dimensional
$N=4$ superconformal non-linear sigma model with $K3$ target space, as in [16,18].
Compactifications of the target space field theory in general yield a `Kaluza-Klein'
tower of massless and massive states. The massless states derived in this way,
together with massless solitons which occur at special moduli values,
are assumed to be the massless spectrum of a generalized 
non-perturbative/perturbative description. Although these techniques do
not provide yet a specification of the massive states which together with
this massless spectrum would serve as a consistent set of states on
which to define the theory.  

On the other hand, computation of the spectrum using the projection operators 
yields both a `Kaluza-Klein' tower of states from the compactification,
and a `Regge' tower of states already present in the $D=10$ IIA string. 
The latter is given in the Appendix (A3). 
This calculation specifies a consistent set of states
for the subset of theories parameterized by $\theta_{\Gamma_{4,4}} (\bar\tau,\tau)$.
That is to say, given the states described in (2.15), there is no need to
add non-perturbative information for the consistency of the theory. 
Nonetheless, the appearance of BPS soliton solutions of the target space 
classical field theory, may be a motivation to modify the
perturbative description to encompass a bigger theory, a modification which
might be essential at larger coupling.  
We  conjecture that a complete definition of the bigger theory will involve
an analog of the perturbative GSO projections.

The massless spectrum of the Type IIA superstring on 
${\bf R}^6\times {{\bf T}^4/{\bf Z}_2}$ is given in terms of representations
of the $D=6$ lightcone little group Spin(4) $\cong SU(2)\times SU(2)$
which form $D=6,\, N = (1,1)$ spacetime supersymmetry multiplets.
The supergravity multiplet is
$$ (3,3) + (3,1) + (1,3) + 4(2,2) + (1,1)
+ 2(3,2) + 2(2,3) + 2(2,1) + 2(1,2)\,.\eqno(3.1a)$$ 
It couples to 20 vector supermultiplets with spin content:
$$ (2,2) + 4(1,1) + 2(2,1) + 2(1,2)\,.\eqno(3.1b)$$
We derive (3.1) in the orbifold model of section 2, using
the projections (2.13) as follows.
The supergravity multiplet and 4 of the vector multiplets come from the
untwisted sector, and 16 vector multiplets come from the twisted sector.
The physical states satisfy
$$ P_{\ag, b_2 }\, P_{\ag, b_0b_2 }\, P_{\ag, b_0b_1}\, |\psi\rangle =
|\psi\rangle \eqno(3.2)$$ 
where
$P_{\ag, b_2 }, P_{\ag, b_0b_2 }, P_{\ag, b_0b_1}$ are defined in (2.13),
via (2.8) and Table 1, and $|\psi\rangle \in \alpha$. 
Writing the four sums in (2.10) as   
$$\eqalignno{\rho_\bg\dt F &= \sum_{j=1}^4\tilde\rho_j \tilde F_j
+ \sum_{j=1}^2\tilde\rho'_j \tilde F'_j
+ \sum_{j=1}^{4}\rho_j F_j
+ \sum_{j=1}^{2}\rho'_j F'_j\cr
&\equiv \bar F_1 + \bar F_2 + \bar F_3 + \bar F_4\,, &(3.3)}$$
we find in the untwisted NS-NS sector ($\alpha=\phi$) that 
(3.2) corresponds to 
$${\textstyle{1\over 8}} [ 1 - (-1)^{\rho_{b_2}\dt (\bar F_1 + \bar F_2)} ] 
[ 1 - (-1)^{\rho_{b_2}\dt (\bar F_3 + \bar F_4)} ] 
[ 1 + (-1)^{(\rho_{b_2}\dt (\bar F_2 + \bar F_4)) + N} ] |\psi\rangle =
|\psi\rangle \eqno(3.4)$$
So for massless states, either $\bar F_1 = \bar F_3 = 0\,;
\bar F_2 = \bar F_4 =1$ or
$\bar F_1 = \bar F_3 = 1\,; 
\bar F_2 = \bar F_4 = 0\,$,
corresponding to $\tilde \psi_{-\hhalf}^I\times \psi_{-\hhalf}^J |0\rangle$
with spin content of 16 scalars
and to $\tilde \psi_{-\hhalf}^\ih\times \psi_{-\hhalf}^\jh |0\rangle$ 
with spin content $(3,3) + (1,3) + (3,1) + (1,1)$.
In the untwisted RR sector $\alpha = b_0$, the projections in (3.2)
on massless states require $\bar F_1 + \bar F_2 =$even;
$\bar F_3 + \bar F_4 =$even, $\bar F_2 + \bar F_4 =$even.
   
The Ramond ground states in $D=6$ for the type IIA superstring 
on ${\bf R}^6\times {\bf T}^4$ or
${\bf R}^6\times {{\bf T}^4/{\bf Z}_2}$ corresponds to the spin content:
$$\eqalignno{\bar F_1 &= {\rm even}\qquad |(2,1)\rangle_{\rm Left}\cr
\bar F_1 &= {\rm odd}\,\,\qquad |(1,2)\rangle_{\rm Left}\cr 
\bar F_3 &= {\rm even}\qquad |(1,2)\rangle_{\rm Right}\cr
\bar F_3 &= {\rm odd}\,\,\qquad |(2,1)\rangle_{\rm Right}\,.&(3.5)}$$ 
Using (3.5), we find the massless states in the untwisted RR sector
have spin content 8(2,2). These eight vectors are from
$2 |(2,1)\rangle_{\rm Left}\times 2|(1,2)\rangle_{\rm Right}$ and
$2|(1,2)\rangle_{\rm Left}\times 2|(2,1)\rangle_{\rm Right}$.
Similar arguments show the R-NS sector contains massless states
$2(2,3) + 10(2,1)$ given by 
$2|(2,1)\rangle_{\rm Left}\times\psi_{-\hhalf}^J |0\rangle$ 
and $2|(1,2)\rangle_{\rm Left}\times\psi_{-\hhalf}^\jh |0\rangle$;
the NS-R sector contains $2(3,2) + 10(1,2)$;
and the four twisted sectors 
contain 16 massless vector supermultiplets given by 
$ 4\, |16\rangle$ and
$|(2,1)\rangle_{\rm Left}\times |(1,2)\rangle_{\rm Right}\times |16\rangle$,
$2\times |(1,2)\rangle_{\rm Right}\times |16\rangle$,
$|(2,1)\rangle_{\rm Left}\times 2 \times |16\rangle$,
where $|16\rangle$ is the degeneracy of the 
ground state of the twisted bosonic operators $\tilde A_s^I, A_s^J$\ref{1}. 

The IIA superstring on $R^6 \times K3$ is conjectured to be S-dual to
the heterotic string on $R^6 \times T^4$\ref{4,5,19}. This requires
both theories to have the same massless spectrum and the same moduli space. 
At points of enhanced symmetry in the moduli space, the massless spectrum
of both theories will be the supegravity multiplet of (3.1a) together
with the vector supermultiplet (3.1b) in the adjoint representation of 
a rank 20 non-abelian group. (For enhanced symmetry, the  
the additional vector supermultiplets imply additional massless scalars,
but these have quartic interactions (as required by the gauge and supersymmetry)
so their vevs are not new moduli.\ref{5}
At generic points in the moduli space, however,
the massless spectrum is given in (3.1); and the 
81 scalar fields can acquire vacuum expectation values
(vevs) which take values in the moduli space ${\cal M} \times {\bf R}$,
where the vev of the dilaton in (3.1a) lives on ${\bf R}$ and 
the vevs of the 80 scalars in the 20 $U(1)$ vector multiplets (3.1b) 
parameterize $${\cal M} = 0(4,20;{\bf Z}) 
\setminus \,{0(4,20)\over{0(4)\times 0(20)}}\,.\eqno(3.6)$$
For the orbifold compactification 
of IIA on ${\bf R}^6\times {{\bf T}^4/{\bf Z}_2}$, 
most of these values have been fixed,
since from (2.15) the partition function depends only on 16 parameters via 
$\theta_{\Gamma_{4,4}}(\bar\tau,\tau)$. 
No choice of these 16 parameters will adjust the $\Gamma_{4,4}$ lattice
to enhance the gauge symmetry, i.e. 
no left or right internal momentum states can appear at the massless level,
since for such states the projections would be $\bar F_1=\bar F_2 =0$
(in sector $\phi$ for example), and this does not satisfy (3.4).  
This reflects the fact\ref{16} that the orbifold which 
leads to a finite string perturbation theory in (2.15)
corresponds to points in the moduli space which do not overlap with
points of enhanced symmetry. 

All 24 of the $U(1)$ vectors in (3.1), i.e. the (2,2)'s of Spin (4), 
are from the RR sectors $b_0$ and $b_1$,
so from conventional arguments\ref{20}
no perturbative states carry any of the 24 $U(1)$ electric charges.    
Also, the 24 magnetic charges are only carried by non-perturbative
states. 

In addition to (3.1),
we compute the lowest lying massive spectrum in perturbation theory.
Expanding (2.15), we see there are $4\dt (64)^2$ at 
mass level $\hhalf m_L^2  = \hhalf m_R^2 = \hhalf$. These
occur in the four twisted sectors, via arguments similar to (3.4),
and have Spin (4) content equivalent to 64 copies of the $D=6$, $N=(2,2)$
supergravity multiplet given in (3.10). Here the states are massive,
with little group Spin (5), and they can be grouped into combinations of 
$D=6$ massive Spin (5) representations (underlined):
$$64 \,[\,{\underline{14}} + 5\;{\underline {10}} + 10\;{\underline {5}}
+ 14\;{\underline {1}}
+  4\;{\underline {16}} + 16\;{\underline {4}} \,]\,.\eqno(3.7)$$
This is 64 unshortened massive supermultiplets, since they do not have the
same spectrum of Spin(4) states as (3.1a). 
(As an example, in sector $b_0b_1$, the states surviving the projections are
$\tilde \psi_{-\hhalf}^\ih |2\rangle 
\times \psi_{-\hhalf}^\jh |2\rangle\times |16\rangle,
\tilde A_{-\hhalf}^I |2\rangle
\times A_{-\hhalf}^J |2\rangle\times |16\rangle,
\tilde \psi_{-\hhalf}^\ih |2\rangle 
\times A_{-\hhalf}^J |2\rangle\times |16\rangle, 
\tilde A_{-\hhalf}^I |2\rangle 
\tilde \psi_{-\hhalf}^\ih |2\rangle$ where $|2\rangle$ refers to the
ground state of the internal integer-moded twisted NS fermions.
These have Spin(4) content 
$64\,[\, (3,3) + (3,1) + (1,3) + 8(2,2) + 17 (1,1)\,]$.
As discussed above, the massive states (3.7) 
are not BPS states (they carry no magnetic charge since they are perturbative,
no electric charge since all gauge fields are RR, and as expected they
do not fit into ultrashort massive supermultiplets). 
They do not couple to the $U(1)$ gauge fields, and thus
their mass presumably receives quantum corrections. It is suggested that
such states may become unstable
from these corrections or non-perturbatively,\ref{5}
because they could decay into BPS states and therefore would not appear in
an exact S-matrix.  
It might be possible there are 
no non-BPS states in the full theory
of IIA on ${\bf R}^6\times {{\bf T}^4/{\bf Z}_2}$, even though they appear
in perturbation theory.\ref{5}

By expanding (A4), we see on the dual side 
(the heterotic string  on $R6\times T4$) that
there are no (perturbative) states at this mass level. 
At the next mass level, $\hhalf m_L^2  = \hhalf m_R^2 = 1$,
there are $64\times 3456$ perturbative states, while 
on the dual side (heterotic), there are 41,472 perturbative states, again demonstrating
the mismatch of perturbative spectra in the S-dual pair.

For the type IIB superstring, the analog of (3.5) is
$$\eqalignno{\bar F_1 &= {\rm even}\qquad |(1,2)\rangle_{\rm Left}\cr
\bar F_1 &= {\rm odd}\,\,\qquad |(2,1)\rangle_{\rm Left}\cr
\bar F_3 &= {\rm even}\qquad |(1,2)\rangle_{\rm Right}\cr
\bar F_3 &= {\rm odd}\,\,\qquad |(2,1)\rangle_{\rm Right}\,.&(3.8)}$$ 
So for the Type IIB superstring on
${\bf R}^6\times {{\bf T}^4/{\bf Z}_2}$
the partition function is the same (2.15) 
as for IIA on ${\bf R}^6\times {{\bf T}^4/{\bf Z}_2}$,
the number of massless states is the same, but the
Spin (4) representations are now form $D=6,\; N=(0,2)$ supermultiplet:
the supergravity multiplet
$$ (3,3) + 5(3,1) + 4(3,2)\eqno(3.9a)$$
is coupled to 21 tensor supermultiplets:
$$ (1,3) + 5(1,1) + 4(1,2)\,.\eqno(3.9b)$$   
From the projections (2.13), the boundary values (2.8), Table 1,
and (3.8), it follows that the massless states in the untwisted sectors are:
$(3,3) + (1,3) + (3,1) + (1,1)$ from NS-NS,
$2 |(1,2)\rangle_{\rm Left}\times 2|(1,2)\rangle_{\rm Right} +
2|(2,1)\rangle_{\rm Left}\times 2|(2,1)\rangle_{\rm Right} =
4(1,3) + 4(3,1) + 8(1,1)$ from RR,
and $4(3,2) + 20(1,2)$ from R-NS and NS-R.
In the twisted sectors, the massless spectrum is
$64(1,1)$ from $b_0b_1$,
$16(1,1) + 16(1,3)$ from $b_1$,
$64(1,2)$ from $b_1b_2$ and $b_0b_1b_2$.
The supergravity multiplet and 5 of the tensor multiplets come from the
untwisted sector, and 16 tensor multiplets come from the twisted sector.     
 
From (3.9), one sees there are no vector multiplets (RR nor NS-NS) in 
Type IIB on ${\bf R}^6\times {{\bf T}^4/{\bf Z}_2}$.
This holds for generic ${\bf R}^6\times {K3}$ compactifications,
so it seems unlikely that special points in the moduli space would lead to 
non-abelian gauge symmetry in $D=6$, unlike the IIA case.

We include here for comparison, the massless spectrum of
type IIA or B on ${\bf R}^6\times {\bf T}^4$. It is 
$D=6,\; N=(2,2)$ with the unique supergravity multiplet:
$$\eqalignno{&(3,3) + 5(3,1) + 5(1.3) + 16(2,2) + 25(1,1)
+ 4(3,2) + 4(2,3) + 20(2,1) + 20(1,2)\cr&&(3.10)}$$ 
where 8 of the vectors come from NS-NS sector, and 8 from RR sector. 
The partition function for this theory is given in the Appendix (A1). 

The lowest lying spectrum for type II on the tube metric conformal field theory
on $R6\times W4/Z_2$ described by the partition function given in (4.4)
for general $k$
is comprised of a $D=6,\; N=(1,1)$  supergravity multiplet and 4
vector supermultiplets,
all in the untwisted sector.

\vskip15pt
\centerline{\bf 4. Partition function for tube metric conformal field theory}
\vskip5pt
The partition function for the type II superstring on ${\bf R}^6\times W4$,
for $W4$ described in sect. 5,  is given by 
$$\eqalignno{\Lambda &= -\,{1\over {4\pi {(\alpha')^3}}}\int_{\cal F}
d^2 \tau (\hbox{\rm Im}\,\tau)^{-2}
\,|\eta(\tau)|^{-8} \,\Lambda'_f\cr
\Lambda'_f &= {\textstyle 1\over 4}
\, {{(\hbox{\rm Im}\,\tau)^{-{1\over 2}}} 
\over |\eta (\tau)|^{2 }}{Z_k(\bar\tau,\tau)}
\quad |\eta (\tau)|^{-8} (\bar\theta_3^4 - \bar\theta_4^4
- \bar\theta_2^4 )  (\theta_3^4 - \theta_4^4 - \theta_2^4)\,,
&(4.1)}$$   
where the diagonal modular invariant 
$$Z_k(\bar\tau,\tau) = 
\sum_{\lambda=1}^{k+1} \bar\chi_{k,\lambda}(\bar\tau)
\chi_{k,\lambda}(\tau) \,\eqno(4.2)$$   
is in correspondence with $su(2+k)$, i.e. 
$A_{k+1}$ in the ADE classification\ref{21}
of modular invariant combinations of level $k$ affine $SU(2)$  characters.
An irreducible highest weight representation of an affine algebra $\hat g$ is an
infinite-dimensional tower of irreducible representations of the
finite-dimensional algebra $g$, and is classified by its highest weight. 
Allowed highest weights for level $k$ affine $SU(2)$ are
$\lambda = 2\ell +1$, where $\ell$ is the spin of the 
$SU(2)$ representation at the top of the tower, 
and $0<\ell<{\textstyle k\over 2}$. 

The character formula, which counts the states in a given irreducible
representation of the level $k$ affine $SU(2)$ is
$$\chi_{k,\lambda}(\tau) = {1\over {\eta^3 (\tau)}}
\sum_{n\in  Z} (n2(k+2) + \lambda)\, 
\omega^{(n2(k+2)+\lambda)^2\over {4(k+2)}}\,.\eqno(4.3)$$
The invariant 
$Z_k(\bar\tau,\tau) =
\sum_{\lambda=1}^{k+1} \bar\chi_{k,\lambda}(\bar\tau)
\chi_{k,\lambda}(\tau)$ is defined for $k\ge 0$. For $k=0$,
$Z_0(\bar\tau,\tau) =
\bar\chi_{0,1}(\bar\tau)
\chi_{0,1}(\tau) = 1$
since $\chi_{0,1}(\tau) = {1\over {\eta^3 (\tau)}}
\omega^{1\over 8}\sum_{n\in  Z} (4n+1)\,
\omega^{2n^2 + n} = 1$ using the Jacobi triple product identity for
$\eta^3$. 

The other ADE modular invariants, for example  
those corresponding to $D_{{k\over 2}+2}$, $k$ even, 
are discussed in [13].

A modular invariant 
partition function for the type II 
superstring on ${\bf R}^6\times W4/{\bf Z}_2$ is 
$$\eqalignno{\Lambda &= -\,{1\over {4\pi {(\alpha')^3}}}\int_{\cal F}
d^2 \tau (\hbox{\rm Im}\,\tau)^{-4}
\,|\eta(\tau)|^{-8} \,\Lambda'_f\cr
\Lambda'_f &= {\textstyle 1\over 8}
\, {{(\hbox{\rm Im}\,\tau)^{-{1\over 2}}}  
\over |\eta (\tau)|^2}\, 
[\,{\ZZ 0. 0.(\bar\tau, \tau)} 
\quad |\eta (\tau)|^{-8} (\bar\theta_3^4 - \bar\theta_4^4
- \bar\theta_2^4 )  (\theta_3^4 - \theta_4^4 - \theta_2^4)\cr
&\hskip60pt +  
{\ZZ 0. 1.(\bar\tau, \tau)}
\quad |\eta (\tau)|^{-8} \,\bar\theta_3^2 \bar\theta_4^2 \theta_3^2
\theta_4^2 \,( 1 - 1 - 1 + 1)\cr
&\hskip60pt +  
{\ZZ 1. 0.(\bar\tau, \tau)}
\quad |\eta (\tau)|^{-8} \,\bar\theta_3^2 \bar\theta_2^2 \theta_3^2
\theta_2^2 \,(2 - 2)\cr
&\hskip60pt +  
{\ZZ 1. 1.(\bar\tau, \tau)} 
\quad |\eta (\tau)|^{-8} \,\bar\theta_4^2 \bar\theta_2^2 \theta_4^2
\theta_2^2 \,(2 - 2)\,\,]
&(4.4)}$$ 
where 
$${\ZZ \alpha. \beta.} (\bar\tau, \tau) =
\sum_{\lambda=1}^{k+1} e^{i\pi\beta (\lambda -1)}
\bar\chi_{k,\lambda}(\bar\tau)\,\;
\chi_{k,\lambda + \alpha (k+2-2\lambda)}\,(\tau)\eqno(4.5)$$
transforms under modular transformations as
$$\eqalignno{
{\ZZ \alpha. \beta.}&\rightarrow
e^{-i\pi ({k\over 2}) \alpha^2} {\ZZ \alpha. \beta +\alpha.} \quad {\rm for}
\quad\tau\rightarrow \tau + 1\cr
{\ZZ \alpha. \beta.}&\rightarrow
e^{i\pi k \alpha\beta} {\ZZ \beta. \alpha.}
\quad\hskip25pt {\rm for} \quad\tau\rightarrow {-{1\over\tau}}
&(4.6)}$$
and
${\ZZ 0. 0.} (\bar\tau, \tau) = Z_k (\bar\tau, \tau)$ in (4.2).
We note that (4.4) is identical to (2.15) when the internal bosons
$A^I$ are replaced with Liouville and WZW modes $J^0,J^i$ (see sect.5).
The ${\bf Z}_2$ twist used in (4.4) is defined in [25]. 

\vskip15pt 
\centerline{\bf 5. The tube metric fivebrane representation }
\vskip5pt

The solitonic fivebrane solution
in sigma-model coordinates has a metric given by
$$ds^2 = \eta_{MN} dx^M dx^N + (1 + (2\pi T_2 )^{-1}\sum_{i=1}^{n_H}
{1\over {(y-y_i)^2}}  )\delta_{mn}
dy^m dy^n\eqno (5.1)$$
where $0\le M,N\le 5$, and $6\le m,n\le 9$, and
magnetic charge $g_6 = {2\pi n_H\over {\sqrt 2} \kappa T_2}$.
The first term corresponds to a $c=9$ flat and therefore free cft, and
the second term represents a non-trivial $c=6$ cft describing $n_H$ coincident
fivebranes when $y_i = 0$.  

For $n_H = k+2$ and near the semi-wormhole throat 
${n_H\over {2\pi T_2 y^2}} \delta_{mn} dy^m dy^n$, 
the $c=6$ cft is an $N=4$ superconformal field theory (scft) whose 
operator algebra 
given in Appendix C has generators constructed from 
four affine Kac-Moody currents $J^A$ of dimension one
satisfying an $U(1)\times SU(2)$ KMA (of level $k$)
together with a set of four dimension-{$\hhalf$} fields $\psi^A$ satisfying
the free fermion algebra:
$$ \eqalignno{J^0 (z) J^0 (\zeta) &= - (z-\zeta)^{-2} + \ldots\cr
J^0 (z) J^i (\zeta) &= O (z-\zeta)^0\cr
J^i (z) J^j (\zeta) &= - {k\over 2} (z-\zeta)^{-2} + \epsilon_{ijk} J^k (\zeta)
(z-\zeta)^{-1} + \ldots\cr
\psi^A (z)\psi^B (\zeta)&= - (z-\zeta)^{-1} \delta^{AB} + \ldots\cr
\psi^A (z) J^B(\zeta)&= O (z-\zeta)^0\,.&(5.2)\cr}$$
The energy-momentum tensor from the Sugawara and Feigen Fuchs constructions is:
$$L(z) = -\half J^0 J^0 - {1\over {k+2}} J^i J^i
-\half\partial\psi^A \psi^A
+\delta {1\over 4 z^2}\,+ \hhalf Q \partial J^0 (z)\eqno (5.3a)$$
where $\delta = 0,1$ in the NS or R sector respectively; and $A = (0,i)$. 
One of the four supercurrents is
$$ F(z) = \psi^0 J^0 + {{\sqrt 2}\over\sqrt{k+2}} \psi^i J^i
+ {{\sqrt 2}\over{6\sqrt{k+2}}} \epsilon_{ijk}\psi^i\psi^j\psi^k
-Q\partial\psi^0 (z)\,,
\eqno(5.3b)$$
and the level 1 $SU(2)$ current in (C1) is constructed as
$$S^i(z) =  \half
(\psi^0\psi^i + \half \epsilon_{ij\ell}\psi^j\psi^\ell)\,.\eqno(5.4)$$
The superconformal system (5.3) has $c = {3k\over {k+2}} +3 + 3 Q^2$.
We choose the background charge $Q = -{{\sqrt 2}\over{\sqrt{k+2}}}$, so
that $c = 6$. From (5.9) we see the Liouville field $J^0(z)=- \partial\phi(z)$
is a space coordinate in this construction, since conventionally
$X^M(z) X^N(\zeta) = - \eta^{MN}
\ln (z-\zeta) + : X^M(z) X^N(\zeta) :$ where the space directions have
$\eta^{ii} = 1$, and hermiticity is defined via
$a^M(z)\equiv i\partial X^M (z) = \sum_n a^{M}_n z^{-n-1}$
as $a^{M\dagger}_n = a^M_{-n}$. 
(Since the Liouville field is a space coordinate in this
application, the Feigen Fuchs shift
in (5.3) is by a real background charge $Q$ which {\it increases} the central
charge by $3Q^2$.)
 
For $k=0$ ($n_H = 2$), the set of fields in (5.2) is reduced to
$J^0(z), \psi^A(z)$ since the superconformal system (5.3) becomes
$$\eqalignno
{&L(z) = -{1\over 2}
J^0 J^0 -{1\over 2}\partial\psi^A \psi^A
+\delta {1\over 4 z^2} - {1\over 2} \partial J^0&(5.5a)\cr
&F(z) = \psi^0 J^0
+ {\textstyle{1\over 6}}
\epsilon_{ijk}\psi^i\psi^j\psi^k + \partial\psi^0\,,&(5.5b)}$$
and the 3 additional supercurrents $F^i(z)= \psi^i J^0 - \psi^0 J^i
-\half\epsilon_{ijk} \psi^0\psi^j\psi^k
+\partial\psi^i$. 

For $k=-1$, the bosonic fields can correspond to the generators of 
a non-compact Wolf space and have positive norm.\ref{22}

For $L(z)$ hermitian, i.e. $L_n^\dagger = L_{-n}$, then
$J_n^{0\dagger} = - J^0_{-n}$ for $n\ne 0$ since Q is real, 
$J_0^{0\dagger} = -Q - J^0_0$ 
$\psi_n^{A\dagger} = -\psi_{-n}^A$, and  $J_n^{i\dagger} = -J^i_{-n}$.
Therefore the states
$J^0_{-n} |\psi\rangle$ for $n>0$ have positive norm:
$$|| J^0_{-n} |\psi\rangle ||^2 = \langle\psi |
(-J^0_n) J^0_{-n} |\psi\rangle =
n \langle \psi |\psi\rangle = n || |\psi\rangle ||^2 > 0\eqno(5.6)$$ for
$||  |\psi\rangle ||^2 > 0$. A similar argument holds for
$\psi^A_{-n} |\psi\rangle$ and $J^i_{-n} |\psi\rangle$.
It follows from the above hermiticity conditions on $J_n^A,\psi_n^A$
that $F_n^\dagger = F_{-n}$.

To study vertex operators consider the primary
weight one-half superfield whose components are
$$V^i_L(z) = \psi^i \,;\qquad V^i_U (z) = Q
(\hhalf \epsilon_{ijk}\psi^j\psi^k +
J^i )\equiv T^i\,,\eqno(5.7a)$$
where
$$\eqalignno{F(z) \psi^i(\zeta)&= (z-\zeta)^{-1}
T^i(\zeta)\cr
F(z) T^i(\zeta) &= (z-\zeta)^{-2} \psi^i(\zeta)
+ (z-\zeta)^{-1}\partial\psi^i(\zeta)\,.&(5.7b)\cr}$$
We define $T^i\equiv  V^i_U$ and (5.7a)
forms a representation of a super Kac-Moody algebra
$$\eqalignno
{&T^i(z) T^j(\zeta) = -{\delta_{ij}\over (z-\zeta)^2} + 
{Q \epsilon_{ijk}T^k(\zeta)\over (z-\zeta)}&(5.8a)\cr
&T^i(z) \psi^j(\zeta) = {Q \epsilon_{ijk} \psi^k(\zeta)\over (z-\zeta)} 
&(5.8b)\cr
&\psi^i(z) \psi^j(\zeta) = -{\delta_{ij}\over (z-\zeta)}\,.&(5.8c)\cr}$$
The level of the $SU(2)$ KMA $T^i$ is $k+2$.
Other relevant $SU(2)$ currents are
$\tilde S^i = \half
(-\psi^0\psi^i + \half \epsilon_{ij\ell}\psi^j\psi^\ell)$ of level 1,
and $\tilde S^i + J^i$ of level $1+k$. 


If we bosonize the $J^0$ current by
$J^0(z)=- \partial\phi(z)$, where
$$\phi (z) \phi(\zeta) = - {\rm ln} (z-\zeta) + \ldots\, 
{\rm for}\, |z|>|\zeta|\,,\eqno(5.9)$$
then the conformal field $:e^{\beta\phi(z)}:$ with
$:e^{\beta\phi(0)}: |0\rangle = |\beta\rangle$   
is primary with respect to $L(z)$
with conformal weight $h_Q = -\hhalf \beta (\beta + Q)$.
We are interested in the case real $Q$ and real $h_Q$. 
The allowed values for $\beta$ such that $h_Q$ is
real are as follows. 
For $h_Q \equiv -\hhalf\beta(\beta + Q)$ to be
real, either $\beta = -\hhalf Q + ip$ with real continuous $p$ so that
$h_Q = {\textstyle{1\over 8}}Q^2 + \hhalf p^2$ is positive;
or $\beta$ is real. If $\beta$ is real, the highest weight conditions
for massless unitary representations
require it to take on discrete values, a feature similar to the discrete
states in $c=1$ matter coupled to $2d$ quantum gravity,
i.e. the Liouville field.\ref{23}

\vskip15pt

\leftline{HERMITICITY PROPERTIES, INNER PRODUCTS AND NORMS}
\vskip5pt  

\def\half{\textstyle{1\over 2}} 

We consider the Liouville part of the cft described in (5.3a) given by\break
${L(z) = -\half:J(z) J(z) : +\half Q\partial J(z)}$, defining 
$J(z)\equiv J^0(z)$ so that
for $n\ne 0$,
$$L_n = - J_0 J_n - {Q\over 2} (n+1) J_n
-\half \sum_{m\ne 0,m\ne n}: J_{n-m} J_m :\eqno(5.10a)$$ and
$$L_0 = -\half J_0^2 - {Q\over 2} J_0
-\sum_{m=1}^\infty: J_{-m} J_m :\eqno(5.10b)$$ 
The current $J(z)$ is anomalous:
$$L(z) J(\zeta) = Q (z-\zeta)^{-3} + J (\zeta) (z-\zeta)^{-2}
+ \partial J (\zeta) (z-\zeta)^{-1} + \ldots \,$$
so that
$$\eqalignno{[L_n, J_m] &= {Q\over 2} n (n+1)\delta_{n,-m} - m J_{n+m}\cr
[L_{-1},  J_1] &= - J_0\cr
[L_{1},  J_{-1}] &= Q + J_0&(5.11)\cr}$$ 
The primary field $e^{\beta\phi(z)}$ satisfies
$$ \eqalignno{L(z) e^{\beta\phi(\zeta)} &=
(z-\zeta)^{-2} [-{1\over 2} \beta(\beta + Q) ]
e^{\beta \phi(\zeta)}
+ (z-\zeta)^{-1}\partial_\zeta e^{\beta\phi(\zeta)} \cr
J(z) e^{\beta\phi(\zeta)}
&= (z-\zeta)^{-1} \beta e^{\beta\phi(\zeta)}\cr
e^{\beta\phi(0)} |0\rangle &= \psi_\beta\cr
J_0 \psi_\beta&=  \beta\psi_\beta&(5.12)\cr}$$
where for $n>0$, $J_n |0\rangle = 0,  L_n |0\rangle = 0$. 
From (5.10a) that for 
$L_n^\dagger = L_{-n}$, we must have
that for $n\ne 0$, $\,J_n^\dagger = -J_{-n}$ if $Q$ is real.
From (5.11) it follows that $J_0^\dagger = - Q - J_0$.  
These hermiticity properties correspond to the following definition of 
inner product: 
$$(\psi_\mu ,\psi_q) = \delta_{\mu, -q-Q}\eqno(5.13)$$
since we must have 
$$\eqalignno{(\psi_\mu , J_0\psi_q) 
&= (J_0^\dagger \psi_\mu ,\psi_q)\cr
= q (\psi_\mu ,\psi_q) &= ((-J_0 - Q) \psi_\mu ,\psi_q)\cr
&= (-\mu - Q) (\psi_\mu ,\psi_q)&(5.14)\cr}$$
so that $(\psi_\mu ,\psi_q) = 0$  unless $ q = -\mu -Q$, hence (5.13).
We can write $(\psi_\mu ,\psi_q) \equiv \langle \mu | q\rangle$  
so that (5.13) is $(\psi_\mu ,\psi_q) = \langle \mu | q\rangle
= \delta_{\mu, -q-Q}\,.$ 
For this definition of hermiticity and corresponding
inner product, we have that the adjoint of $|q\rangle$ is
$\langle q |$ and that
$$(\psi_q ,\psi_q) = \delta_{q, -q-Q} =  \delta_{q, -{Q\over 2} }\eqno(5.15)$$ i.e.
$\psi_q \equiv |q\rangle$ has zero norm unless $q = -{Q\over 2}$.
Note however that $|q\rangle$ is not null, i.e. 
it is not orthogonal to every state, since
$(\psi_{-q-Q} ,\psi_q) = \delta_{q, q} = 1 \ne 0$.
If the eigenvalues of $J_0$ are complex, then the RHS of (5.14) is 
$(-\mu^\ast -Q) (\psi_\mu ,\psi_q)$, so that (5.15) is
$$(\psi_q ,\psi_q) = \delta_{q, -q^\ast -Q} =  \delta_{{\rm Re} q, -{Q\over 2} }$$ i.e.
$\psi_q \equiv |q\rangle$ has {\bf zero norm unless} 
$${\rm Re} q = -{Q\over 2}\,.\eqno(5.16)$$
(Of course we could construct a positive norm state
$\psi = 
{\scriptstyle {1\over \sqrt 2}} \,(|q\rangle + |-q-Q\rangle)$
where $||\psi ||^2 = 1$. But then there would also be negative norm state
$\psi' = 
{\scriptstyle {1\over \sqrt 2}} \,(|q\rangle - |-q-Q\rangle)$
where $||\psi' ||^2 = -1$, unless this were absent by projection.)
In sect.4, we have assumed the definition of inner product (5.13),
so that corresponding to (5.16),
positive norm states have $J_0$ eigenvalues $\beta = -{Q\over 2} + ip$,
with $h_Q = {\textstyle{1\over 8}}Q^2 + \hhalf p^2$. Integrating over $p$
we compute the contribution of the non-compact Liouville mode
${{(\hbox{\rm Im}\,\tau)^{-{1\over 2}}}
\over |\eta (\tau)|^2}$ appearing as the first factor in 
the partition functions (4.1),(4.4). (Although these theories have 
this (seventh) non-compact dimension, there is only six-dimensional 
Lorentz invariance.)

\vskip 5pt
Let's now consider a {\bf new} inner product
$$\langle\langle \psi_\mu ,\psi_q \rangle\rangle = \delta_{\mu, q}.\eqno(5.17)$$
 This corresponds to the hermiticity properties 
$J_0^\dagger = J_0$, and for $n\ne 0, \,J_n^\dagger = - J_{-n}$ if Q is real,
which results in $L_n^\dagger = L_{-n} + (2J_0 + Q) J_{-n}$ for $n\ne 0$, and
$L_0^\dagger = L_0$. (In a supersymmetric extension 
$F_r^\dagger = F_{-r} -(2J_0 + Q)\psi_{-r}^0.$\ref{24}
To show $J_0^\dagger = J_0$, we have from (5.17) that
$$\eqalignno{\langle\langle \psi_\mu , J_0\psi_q \rangle\rangle &=
\langle\langle J_0^\dagger \psi_\mu , \psi_q \rangle\rangle&(5.18a)\cr
= q \langle\langle \psi_\mu ,\psi_q \rangle\rangle 
&= \mu \langle\langle \psi_\mu ,\psi_q \rangle\rangle&(5.18b)\cr
&= \langle\langle J_0\psi_\mu , \psi_q \rangle\rangle&(5.18c)\cr}$$
where (5.18b) follows from 
$\langle\langle \psi_\mu ,\psi_q \rangle\rangle = \delta_{\mu, q}$,
and $J_0^\dagger = J_0$ follows from (5.18a,c).
Then$$\eqalignno{\langle\langle \psi_\mu ,\psi_q \rangle\rangle
&\equiv (\psi_{-\mu -Q},\psi_q) = \delta_{\mu, q}\cr
\langle\langle \psi_\mu ,\psi_q \rangle\rangle  
&\equiv \langle -\mu -Q| q\rangle
= \delta_{\mu, q}\,.&(5.19)\cr}$$ 
For this definition of hermiticity and corresponding
inner product, we have that the adjoint of $|q\rangle$ is
$\langle -q-Q |$ and that
$$\langle\langle\psi_q ,\psi_q\rangle\rangle 
= \delta_{q, q} = 1\,,\eqno(5.20)$$ i.e.
$\psi_q \equiv |q\rangle$ has unit norm.
If the eigenvalues of $J_0$ are complex, then the RHS of (5.18b) is
$\mu^\ast \langle\langle \psi_\mu ,\psi_q \rangle\rangle$, so that (5.20) is
$\langle\langle\psi_q ,\psi_q\rangle\rangle = \delta_{q^\ast, q} = 1\,,$ i.e.
$\psi_q \equiv |q\rangle$ has {\bf non-zero norm} for any real q.
With the new inner product, the discrete states have positive norm. 
This new definition of norm is seen to agree with that found in recent exact
results on Liouville theory\ref{17}. Incorporation of these results as
internal degrees of freedom of a string theory would allow us to remove
the non-compact contribution ${{(\hbox{\rm Im}\,\tau)^{-{1\over 2}}}
\over |\eta (\tau)|^2}$, which is modular invariant by itself.  

\vskip15pt

\leftline{D=2 SUPERSYMMETRIC GAUGE THEORIES}

Following recent work on two-dimensional theories, we recall\ref{11-13}
that $D=2$, $N=(4,4)$ supersymmetric gauge theories can have
vector supermultiplets $[\,2(\pm\half), 4(0)\,]$ with gauge group G,
where the scalars are
labelled by $\Phi = \phi_i^a$,$\,$ $1\le i\le 4; 1\le a\le {\rm dim}\,G$,
and hypermultiplets $[\,2(\pm\half), 4(0)\,]$ with scalars
$H^{AX}$,$\,$ $1\le A\le 2; 1\le X\le 2n_H$, where $n_H$ is 
the number of hypermultiplets. (These $2D$ multiplets are dimensionally reduced
$D=6$, $N=(1,0)$ vector $[\,(2,2) + 2(1,2)\,]$ and hypermultiplets 
$[\,2(2,1) + 4(1,1)\,]$, here labelled by their Spin(4) content.) 
A Coulomb phase has
$\langle 0 | H | 0\rangle = 0;\, \langle 0 | \Phi | 0\rangle\ne 0$ and
the surviving massless fields are $r$ massless vector supermultiplets, 
where r is the rank of $G$.
The infrared limit
of this $D=2$ supersymmetric gauge theory is an $N=(4,4)$ superconformal
field theory with central charge $c = 6r$. For $r=1$ it 
corresponds to the tube metric cft which has primary fields 
(4 scalars (bosons) and 4 fermions) given in (5.2). 
The level of the $SU(2)$ KMA $J^i$ in (5.2) is $k = n_H -2$, so a 2D gauge theory with 
$n_H$ hypermultiplets flows in the infrared 
to a scft corresponding to the 
transverse degrees of freedom to $n_H$ coincident fivebranes (5.1).
In IIB for example, the 2D gauge theory is the world volume theory of a D1-string 
containing the $U(1)$ gauge field which can carry away the flux associated with
$n_H$ point electric charges, each emerging from the end of a fundamental open 
string connected to one of $n_H$ nearby D5-branes.
(The world volume action of $n_H$ coincident Dirichlet fivebranes is
a $D=6$, N=(1,1), $U(n_H)$ non-abelian gauge theory). 
These become Neveu-Schwarz fivebranes in the S-dual picture.

The modular invariant character combinations for level $k$ $\widehat{SU}(2)$
were found by Cappelli, Itzykson, Zuber\ref{21} to have an ADE classification
$A_{k+1},\, k\ge 0;\; D_{{k\over2}+2},\, k\ge 4, k {\rm even};\;
E_6,\, k=10;\, E_7,\, k=16\;;E_8,\, k=28$. 
The $D=2$ gauge theory with an $SU(2)$ vector multiplet and $n_H$ hypermultiplets
corresponds to the $D_{{k\over2}+2}$ series\ref{13}.
The $A_{k+1}$ modular invariants correspond to one $U(1)$ vector multiplet
and $n_H = k+2$ hypermultiplets; for a given $k$, these invariants are 
$Z_k(\bar\tau,\tau) =
\sum_{\lambda=1}^{k+1} \bar\chi_{k,\lambda}(\bar\tau)
\chi_{k,\lambda}(\tau)$ described in (4.2), where 
$\chi_{k,\lambda}(\tau)$ is the character for the highest weight 
$\lambda$ irreducible representation of the level $k$
affine $SU(2)$ KMA.   

The Higgs phase of the 2D theory has $\langle 0 | H | 0\rangle \ne 0;\, 
\langle 0 | \Phi | 0\rangle = 0$ and is parameterized by 
$n_H- n_V$ massless hypermultiplets, with infrared limit of
an $N=(4,4)$ superconformal
field theory with central charge $c = 6(n_H - n_V)$.

In the dual pair: heterotic on $R^6\times T^4$ and type IIA on
$R^6\times K3$, the enhanced symmetry points 
in the $K3$ moduli space (3.6) have an ADE classification
corresponding to simply laced non-abelian symmetries of perturbative states
in the heterotic theory. Thus to study a point in this moduli space corresponding
to $A_1\cong SU(2)$ i.e., a theory whose massless spectrum in spacetime 
is a $D=6$, $N=(1,1)$ supergravity multiplet coupled to 19 $U(1)$ and
1 $SU(2)$ vector multiplets, we look on the type II side at the IIA theory which 
is T-dual to the particular IIB compactification described by 
the $2D$ gauge theory $N=(4,4)$ with $U(1)$ vector multiplet ($n_V=1$) and 
$n_H= k+2 = 2$ hypermultiplets; this in turn corresponds to the $k=0$ tube metric
cft generated by (5.5) with primary fields $J^0(z), \psi^0(z)$ and $\psi^i(z)$
(on the left and right-moving sides), to describe the 
non-spacetime degrees of freedom.  

\vskip15pt
\centerline{\bf 6. Conclusions}
\vskip5pt
\parskip=6pt

The partition functions (4.1) and (4.4) are constructed from the
$A_{k+1}$ modular invariant $Z_k(\bar\tau,\tau)$ and related twisted expressions
(4.5). These correspond to excitations of a type II 
fundamental string in a background
described by degrees of freedom transverse to the NS fivebrane.
For general $k$, the lowest lying spectrum is 
a $D=6$ supergravity multiplet and 4 $U(1)$ vector multiplets coming from
the untwisted sector. (For $k=0$ the compatibility of this $Z_2$ twist
with the global existence of the 2D supercurrent (5.3b),
(5.5b) remains problematic\ref{25}.) The incorporation of exact results on
Liouville cft may modify which states survive in this theory, and hence their
gauge symmetry properties. We note the occurrence of the $A_{k+1}$ modular 
invariants, and contrast this theory   
with a type II compactification\ref{11-13}
described by a 2D supersymmetric gauge theory leading to  
a $D=6$, $N=(1,1)$ theory with massless spectrum of
1 SG multiplet, 19 $U(1)$ and 1 $SU(2)$ vector supermultiplets.
A deeper understanding of how the conventional
type II cft breaks down in this case, due to massless solitons,
may guide us to a more economical description of how string 
theory picks the vacuum. 
It is believed that the appearance of these massless non-perturbative BPS states 
may be an important mechanism for the way in which nature incorporates gauge
symmetry in string theory. 

Of course, some non-abelian symmetry can be incorporated in conventional 
type II perturbation theory\ref{26}:
in Appendix B, we give a model whose
internal cft is constructed from free fermions, and which has
a $6D$, $N=(1,1)$ massless spectrum with 49 scalar fields, and
$(SU(2))^4$ non-abelian symmetry. This latter has a 17-dimensional
moduli space corresponding to the 16 massless scalars in
the $U(1)^4$ Cartan subalgebra vector supermultiplets, and the dilaton,
and so is on a different moduli space of vacua. 

\vskip20pt

\leftline{\it Acknowledgements} We acknowledge valuable conversations with
P. Aspinwall, D.-E. Diaconescu, P. Goddard, M. Green, K. Intriligator, I. Klebanov,
E. Martinec, D. Morrison, R. Plesser, J. Schwarz and E. Witten.
L.D. thanks the Aspen Center for Physics for its hospitality. 

\vfill\eject
\centerline{\bf Appendix A} 

\vskip10pt
\leftline{ONE-LOOP PARTITION FUNCTIONS FOR $D=6$ STRING THEORIES:}
\noindent In addition to expressions for type II on the orbifold 
and fivebrane given in sections 2,4, we list here other backgrounds 
for type II and heterotic strings.

\vskip10pt
\leftline{Type II superstring on  ${\bf R}^6\times {\bf T}^4:$}

\vskip-20pt
$$\eqalignno{\Lambda &= -\,{1\over {4\pi {(\alpha')^3}}}\int_{\cal F}
d^2 \tau (\hbox{\rm Im}\,\tau)^{-4}
\,|\eta(\tau)|^{-8} \,\Lambda'_f\cr
\Lambda'_f &= {\textstyle 1\over 4}
\, {\theta_{\Gamma_{4,4}} (\bar\tau,\tau)\over |\eta (\tau)|^8 }
\quad |\eta (\tau)|^{-8} (\bar\theta_3^4 - \bar\theta_4^4
- \bar\theta_2^4 )  (\theta_3^4 - \theta_4^4 - \theta_2^4)\,.
&(A1)}$$
\vskip15pt
\leftline{Type II superstring on  ${\bf R}^6\times {\bf R}^4/{\bf Z}_2:$} 

\vskip-20pt
$$\eqalignno{\Lambda &= -\,{1\over {4\pi {(\alpha')^3}}}\int_{\cal F}
d^2 \tau (\hbox{\rm Im}\,\tau)^{-4}
\,|\eta(\tau)|^{-8} \,\Lambda'_f\cr
\Lambda'_f &= {\textstyle 1\over 8}
[\, {(\hbox{\rm Im}\,\tau)^{-2}\over |\eta (\tau)|^8 } 
\quad |\eta (\tau)|^{-8} (\bar\theta_3^4 - \bar\theta_4^4
- \bar\theta_2^4 )  (\theta_3^4 - \theta_4^4 - \theta_2^4)\cr
&\hskip20pt +  {2^4 |\eta (\tau)|^4 \over \bar\theta_2^2 \theta_2^2}
\quad |\eta (\tau)|^{-8} \,\bar\theta_3^2 \bar\theta_4^2 \theta_3^2
\theta_4^2 \,( 1 - 1 - 1 + 1)\cr
&\hskip20pt +  {2^4 |\eta (\tau)|^4 \over \bar\theta_4^2 \theta_4^2}
\quad |\eta (\tau)|^{-8} \,\bar\theta_3^2 \bar\theta_2^2 \theta_3^2
\theta_2^2 \,(2 - 2)\cr
&\hskip20pt +  {2^4 |\eta (\tau)|^4 \over \bar\theta_3^2 \theta_3^2}
\quad |\eta (\tau)|^{-8} \,\bar\theta_4^2 \bar\theta_2^2 \theta_4^2
\theta_2^2 \,(2 - 2)\,\,]\,.
&(A2)}$$ 
\vskip15pt
\leftline{Type II superstring on  ${\bf R}^{10}:$}

\vskip-20pt
$$\eqalignno{\Lambda &= -\,{1\over {4\pi {(\alpha')^3}}}\int_{\cal F}
d^2 \tau (\hbox{\rm Im}\,\tau)^{-4}
\,|\eta(\tau)|^{-8} \,\Lambda'_f\cr
\Lambda'_f &= {\textstyle 1\over 4}
\, {(\hbox{\rm Im}\,\tau)^{-2}  \over |\eta (\tau)|^8 }
\quad |\eta (\tau)|^{-8} (\bar\theta_3^4 - \bar\theta_4^4
- \bar\theta_2^4 )  (\theta_3^4 - \theta_4^4 - \theta_2^4)\,.
&(A3)}$$       
\vfill\eject 
\leftline{Heterotic string on  ${\bf R}^6\times {\bf T}^4:$}

\vskip-20pt
$$\eqalignno{\Lambda &= -\,{1\over {4\pi {(\alpha')^3}}}\int_{\cal F}
d^2 \tau (\hbox{\rm Im}\,\tau)^{-4}
\,|\eta(\tau)|^{-8} \,\Lambda'_f\cr
\Lambda'_f &= {\textstyle 1\over 2}
\, {\theta_{\Gamma_{20,4}} (\bar\tau,\tau)\over \eta (\bar\tau)^{-20}}
\quad |\eta (\tau)|^{-8} (\theta_3^4 - \theta_4^4 - \theta_2^4)\,. 
&(A4)}$$
\vskip15pt 
\leftline{Heterotic string on  ${\bf R}^{10}:$}                                                               
\vskip-20pt
$$\eqalignno{\Lambda &= -\,{1\over {4\pi {(\alpha')^3}}}\int_{\cal F}
d^2 \tau (\hbox{\rm Im}\,\tau)^{-4}
\,|\eta(\tau)|^{-8} \,\Lambda'_f\cr
\Lambda'_f &= {\textstyle 1\over 2}
\,  {(\hbox{\rm Im}\,\tau)^{-2}  \over |\eta (\tau)|^8 } \,
{\theta_{\Gamma_{16}} (\bar\tau)\over \eta (\bar\tau)^{16}}
\quad \eta (\tau)^{-4} (\theta_3^4 - \theta_4^4 - \theta_2^4)\,.
&(A5)}$$
where the lattice theta function 
$$\theta_{\Gamma_{16}} (\bar\tau) =
{\textstyle 1\over 2} (\theta_3^{16} + \theta_4^{16} + \theta_2^{16})$$
is identical for the two 16-dimensional even self-dual Euclidean lattices. 

\vskip15pt
\vfill\eject
\centerline{\bf Appendix B}
\vskip10pt
\leftline{$D=6$, $N=(1,1)$  FREE FERMION MODEL}

A type II string free fermion model with a $D=6$, $N=(1,1)$ massless spectrum
of one supergravity multiplet from $\tilde \psi_{-\hhalf}^\ih\times 
\psi_{-\hhalf}^\jh |0\rangle\; , \tilde \psi_{-\hhalf}^\ih\times
\psi_{-\hhalf}^J |0\rangle\;, \tilde \psi_{-\hhalf}^\ih\times
2|(2,1)+(1,2)\rangle_{\rm Right}\;$;
and 12 vector supermultiplets in the adjoint of $SU(2)^4$
from $\tilde \psi_{-\hhalf}^a\times \{\psi_{-\hhalf}^\jh |0\rangle\,, 
\psi_{-\hhalf}^J |0\rangle\ , 2|(2,1)+(1,2)\rangle_{\rm Right}\}$,
has left and right-moving modes each described by
4 bosonic and 12 fermionic  worldsheet (primary)
fields:
$\tilde A^\ih_n, \tilde \psi_r^\ih, \tilde \psi_r^a$;
$A^\ih_n,\psi_r^\ih, \psi_r^J, \psi_r^j$; $1\leq \ih\leq 4$, 
$1\leq a\leq 12$,
$1\leq J\leq 4$, and $1\leq j\leq 8$ where the
superscript $\ih$ refers to the transverse spatial degrees of freedom,
the superscripts $a,J,j$ to the internal ones, and all bosons and fermions are real.
This string has four sectors
whose fermion boundary condition vectors are given by
$$\eqalignno{\rho_{b_1}&= (0^{16}; 1^4 0^{12})\hskip40pt
\rho_{b_0}= (1^{16}; 1^{16})\cr 
\rho_{b_0b_1}&= (1^{16}; 0^4 1^{12})\hskip42pt
\rho_{\phi}= (0^{16}; 0^{16}) \,.&(B1)}$$ 
The projections replacing (2.13) are
$$\eqalignno{P_{\ag, b_1 } &= \half \{1 + \epsilon(\ag,b_1)
(-1)^{\rho_{b_1}\dt F}\}&(B2)\cr
P_{\ag, b_0b_1 } &= \half \{1 + \epsilon(\ag,b_0b_1)
(-1)^{\rho_{b_0b_1}\dt F}\}\,&(B3)}$$  
where $\rho_\beta\dt F = \sum_{j=1}^{16}\tilde\rho_j \tilde F_j
+ \sum_{j=1}^{16}\rho_j F_j$,
and the values of $\epsilon(\alpha,\beta)$ are found in Table 2.

\hfil\hbox{\raise 15mm\vbox{\hbox{$\ag$}\hbox{\big\downarrow}}
\vbox{\offinterlineskip
\halign{\vrule#&
  \strut\quad\hfil#\quad&\vrule#&
  \strut\quad\hfil#\quad&
  \strut\quad\hfil#\quad&
  \strut\quad\hfil#\quad&
  \strut\quad\hfil#\quad&\vrule#\cr
\multispan8 \hfil $\bg \longrightarrow $\hfil\cr
\noalign{\vskip 4pt\hrule}
height2pt&\omit&&\omit&\omit&\omit&\omit&\cr
&$\epsilon(\ag,\bg)$\hfil&&\O&$b_0$&$b_1$&$b_0b_1$&\cr
height2pt&\omit&&\omit&\omit&\omit&\omit&\cr
\noalign{\hrule}
height2pt&\omit&&\omit&\omit&\omit&\omit&\cr
&\O&&1&1&$-1$&$-1$&\cr  
&$b_0$&&1&$\lambda\mu$&$\mu$&$\lambda$&\cr
&$b_1$&&1&$-\mu$&$\mu$&$-1$&\cr
&$b_0b_1$&&1&$-\lambda$&$-1$&$\lambda$&\cr
height2pt&\omit&&\omit&\omit&\omit&\omit&\cr
\noalign{\hrule\vskip8pt }
\multispan8 \hfil {\sl Table 2}\hfil\cr
}}}\hfil
\vskip6pt  
The partition function is
$$\eqalignno{\Lambda &= -\,{1\over {4\pi {(\alpha')^3}}}\int_{\cal F}
d^2 \tau (\hbox{\rm Im}\,\tau)^{-4}
\,|\eta(\tau)|^{-8} \,\Lambda'_f&(B4)\cr
\Lambda'_f &= {\textstyle 1\over 4}
\quad |\eta (\tau)|^{-16} (|\theta_3|^8\, \bar\theta_3^4 - 
|\theta_4|^8 \,\bar\theta_4^4
- |\theta_2|^8 \,\bar\theta_2^4 )  (\theta_3^4 - \theta_4^4 - \theta_2^4)\,.
&(B5)}$$

\vfill\eject
\centerline{\bf Appendix C}
\vskip10pt
\leftline{$N=4$, $c=6$ SUPERCONFORMAL ALGEBRA}

The currents of (2.5) close on the $N=4$ superconformal algebra
with $c=6$ which is given by 
$$\eqalignno{&L(z) L(\zeta) = {{c\over 2}\over (z-\zeta)^4}
+ {2L(\zeta)\over{(z-\zeta)^2}} + {\partial L(\zeta)\over (z-\zeta)}\cr
&L(z) F(\zeta) = {{3\over 2}F(\zeta)\over (z-\zeta)^2} +
{\partial F(\zeta)\over(z-\zeta)}\cr
&L(z) F^i(\zeta) = {{3\over 2}F^i(\zeta)\over (z-\zeta)^2} + 
{\partial F^i(\zeta)\over (z-\zeta)}\cr
&L(z) S^i(\zeta) = {S^i(\zeta)\over (z-\zeta)^2} +
{\partial S^i(\zeta)\over (z-\zeta)}\cr
&F(z) F(\zeta) = {{2c\over 3}\over (z-\zeta)^3} +
{2L(\zeta)\over{(z-\zeta)}}\cr
&F(z) F^i(\zeta) = - {4 S^i(\zeta)\over (z-\zeta)^2} 
-  {2\partial S^i(\zeta)\over{(z-\zeta)}}\cr
&F^i(z) F^j(\zeta) = {\delta^{ij}{2c\over 3}\over (z-\zeta)^3}
- {4 \epsilon_{ij\ell} S^\ell(\zeta)\over (z-\zeta)^2} 
- {2 \epsilon_{ij\ell} \partial S^\ell (\zeta)\over{(z-\zeta)}}  
+ {2 \delta^{ij} L(\zeta)\over{(z-\zeta)}}\cr
&S^i(z) S^j(\zeta) = -{n\delta^{ij}\over 2 (z-\zeta)^2}            
+ {\epsilon_{ij\ell} S^\ell(\zeta)\over (z-\zeta)}\cr 
&S^i(z) F(\zeta) =  { F^i(\zeta)\over 2 (z-\zeta)}\cr
&S^i(z) F^j(\zeta) = {1\over (z-\zeta)} 
[ -\delta^{ij} F(\zeta) + \epsilon_{ij\ell} F^\ell (\zeta) ]\,.&(C1)\cr}$$
The central charge $c$ and the level $n$ of the $SU(2)_n$ currents
$S^i$ are related by $c = 6n$. The condition $c=6$ sets
$S^i$ at level one.  

We define complex supercurrents as
$$\eqalignno{{\cal G}^1 &\equiv {F - i F^3\over \sqrt 2}\,,\qquad
{\cal G}^2 \equiv {F^2 - i F^1\over \sqrt 2 }\cr
\bar{\cal G}^1 &\equiv {F + i F^3\over \sqrt 2}\,,\qquad
\bar{\cal G}^2 \equiv {F^2 + i F^1\over \sqrt 2}\cr}$$
and note that ${\cal G}^1,{\cal G}^2$ transform as a doublet under
$S^i$ etc.

\vfill\eject
\centerline{\bf References}
\vskip 5pt
\item{ 1.} R. Bluhm, L. Dolan and P. Goddard, 
``Unitarity and Modular Invariance
as Constraints on Four-dimensional Superstrings'',
Nucl. Phys. {\bf B309} (1988) 330. 
\item{ 2.} I. Antoniadis, C. Bachas and C. Kounnas, Nucl. Phys.
{\bf B289} (1987) 87. 
\item{ 3.} M.B. Green, J. Schwarz and E. Witten, {\it Superstring
Theory, Volume 1}, Section 7.3 (Cambridge University Press, 1987).
\item{ 4.} E. Witten, ``String theory in various dimensions'',
Nucl. Phys. {\bf B443} 85 (1995), hep-th/9503124.  
\item{ 5.} C. Hull and P. Townsend, Nucl. Phys. {\bf B438} 109 (1995),
hep-th/9410167; Nucl. Phys. {\bf B451} 525 (1995),
hep-th/9505073.
\item{ 6.} T. Banks, W. Fischler, S. Shenker, and L. Susskind,
Phys. Rev. {\bf D55} (1996) 5112,  hep-th/9610043;
T. Banks, ``Matrix Theory'', hep-th/9710231.
\item{ 7.}  A. Strominger,
``Heterotic Solitons'', Nucl. Phys. {\bf B343} (1990) 167.
\item{ 8.} C. Callan, J. Harvey, and A. Strominger,
``Worldsheet approach to heterotic
instantons and solitons'', Nucl. Phys. {\bf B359} (1991) 611; and
``Worldbrane actions for string solitons'', Nucl. Phys.
{\bf B 367} (1991) 60.
\item{ 9.} M.Duff, R. Khuri, and J.X. Lu, ``String Solitions'',
hep-th/9412184 (Phys. Rep.), and references therein. 
\item{ 10.} C. Callan, ``Instantons and Solitons in Heterotic String Theory'',
Swieca Summer School, June 1991; hep-th/9109052.
C. Callan, J. Harvey, and A. Strominger, ``Supersymmetric String Solitions'',
1991 Trieste Spring School on String Theory and Quantum Gravity,
hep-th/9111030. 
\item{11.} E. Witten, ``Some Comments On String Dynamics'',
{\it Strings '95} (World Scientific, 1996),
ed. I. Bars et. al., 501, hep-th/9507021.
\item{12.} E. Witten, ``On The Conformal Field Theory Of The Higgs Branch'',
hep-th/9707093.  
\item{13.} D. Diaconescu and N. Seiberg, ``The Coulomb Branch of (4,4)
Supersymmetric Field Theories In Two Dimensions'', hep-th/9707158.   
\item{14.} E. Witten, ``New Gauge Theories in Six Dimensions'',
hep-th/9710065. 
\item{15.} P. Aspinwall, ``Enhanced Gauge Symmetries And K3 Surfaces'',
hep-th/9507012, Phys. Lett. {\bf B357} (1995) 329.   
\item{16.} P. Aspinwall. ``K3 Surfaces And String Duality'',
hep-th/9611137. 
\item{17.} A.B. Zamolodchikov and Al.B. Zamolodchikov,
``Structure Constants and Conformal Bootstrap in Liouville Field Theory'',
hep-th/9506136.
\item{18.} J. Schwarz, ``Lectures On Superstring And M Theory Dualities'',
1996 TASI Lectures, Nucl. Phys. Proc. Suppl.
{\bf 55B}: 1-32 (1997), hep-th/9607201. 
\item{19.} P. Aspinwall and D. Morrison, ``String Theory On K3 Surfaces'',
hep-th/9404151, {\it Mirror Symmetry II}, Greene, B. and Yau, S., eds.  
\item{20.} L. Dixon, V. Kaplunovsky and C. Vafa, Nucl. Phys.
{\bf B294} (1987) 43. 
\item{21.} A. Cappelli, C. Itzykson, J.B.Zuber,
Nucl. Phys. {\bf B280} (1987) 445; Comm. Math. Phys.
{\bf 113} (1987) 1.  
\item{22.} Ph. Spindel, A. Sevrin, W. Troost, and A. Van Proeyen,
``Extended Supersymmetric $\sigma$-Models on Group Manifolds'',
Nucl. Phys. {\bf 308} (1988) 662; {\bf B311} (1988) 465. 
\item{23.} I. Klebanov and A. Polyakov, Mod. Phys. Lett A6(1991)635,
hep-th/9109032;
I. Klebanov and A Pasquinucci, hep-th/9210105.  
\item{24.}L. Dolan, ``Gauge Symmetry in Background Charge Conformal Field Theory'',
Nucl. Phys.{\bf B489} (1997) 245, hep-th/9610091. 
\item{25.} I. Antoniadis, S. Ferrara, C. Kounnas,
``Exact supersymmetric string
solutions in curved gravitational backgrounds'',
Nucl. Phys.{\bf B421} (1994) 343, hep-th/9402073. 
\item{26.} R. Bluhm, L. Dolan and P. Goddard, ``A New Method of Incorporating
Symmetry in Superstring Theory'', Nucl. Phys.{\bf B289} 364 (1987). 


\vfill\eject
 
\bye